\documentclass[11pt]{article}
\usepackage{moriond,epsfig}

\def\Journal#1#2#3#4{{#1} {\bf #2}, #3 (#4)}

\def\NIMA{{\em Nucl. Instrum. Methods} A}
\def\NPB{{\em Nucl. Phys.} B}
\def\PLB{{\em Phys. Lett.}  B}

\def\PRD{{\em Phys. Rev.} D}

\def\be{\begin{equation}}
\def\ee{\end{equation}}
\def\bea{\begin{eqnarray}}
\def\eea{\end{eqnarray}}

\begin{document}
\vspace*{4cm}
\title{Spin dependent fragmentation functions at Belle}

\author{ R.Seidl}
\address{University of Illinois at Urbana-Champaign,Department of Physics,\\ 1110 W Green St., 61801 Urbana,Il, USA}

\maketitle\abstracts{Spin dependent fragmentation functions are an important tool to understand the spin structure of the nucleon. The Collins fragmentation function for example describes the fragmentation of a transversely polarized quark into a hadron via the creation of additional transverse momentum relative to the quark's momentum. The Belle experiment has measured the resulting azimuthal asymmetries in a data sample of 29 fb$^{-1}$ and found significant nonzero results.
}

\section{Introduction}
The transverse spin of quarks in the nucleon has so far not been measured although it is also one of the three leading twist quark distribution functions (DF) in the nucleon. The others are the quark helicity distribution and the number density. The reason for this is that transversity is a chiral-odd DF and thus is not accessible in deep inelastic scattering (DIS). When combining it with a second chiral-odd function, the whole process becomes chiral-even again and one can measure it for example in transversely polarized Drell-Yan experiments or semi-inclusive DIS (SIDIS). In the latter the chiral-odd function will be a fragmentation function (FF) such as the Collins function \cite{collins} or the interference fragmentation function \cite{jaffe}. Both of these functions will create azimuthal asymmetries of either one or two hadrons around the momentum of the quark, where the angle is defined relative to the quark's spin. Obviously, these two FFs are also interesting in connecting the microscopic spin of the quark to an macroscopic observable through nonperturbative QCD processes. 
Since the SIDIS experiments are dealing with two unknown functions it is important to obtain the fragmentation functions from $e^+e^-$ annihilation. At the Belle experiment one has the opportunity to obtain these from a high statistics data sample.
\section{The Belle experiment}
The present studies were performed using a $29\ fb^{-1}$ data sample collected at a center of mass energy (CMS) 60 MeV below the $\Upsilon(4S)$ resonance with the Belle detector operating at the KEKB asymmetric energy $e^+e^-$ storage rings \cite{kekb}. The detector is a large-solid-angle magnetic spectrometer that
consists of a multi-layer silicon vertex detector (SVD), a 50-layer central drift chamber (CDC), an array of aerogel threshold \v{C}erenkov counters (ACC), a barrel-like arrangement of time-of-flight scintillation counters (TOF), and an electromagnetic calorimeter (ECL) comprised of CsI(Tl) crystals located inside a superconducting solenoid coil that provides a $1.5$ T magnetic field. An iron flux-return located outside of the coil is instrumented to detect $K^0_L$ mesons and
to identify muons (KLM). The detector is described in detail elsewhere \cite{belledetector}. Two different inner detector configurations were used. For the first data sample of 15.8 $fb^{-1}$
, a $2.0$ cm radius beampipe and a 3-layer silicon vertex detector were used; for the latter 13.2 $fb^{-1}$
, a $1.5$ cm radius beampipe, a 4-layer silicon detector and a small-cell inner drift chamber were used.
 For systematic checks Monte Carlo (MC) simulated events were used generated by the QQ generator and JETSET \cite{jetset} and processed with a full simulation of the Belle detector using the GEANT package \cite{geant}.
\paragraph{Collins Effect}
The Collins effect occurs in the fragmentation of a transversely polarized quark with polarization $\mathbf{S_q}$ and 3-momentum $\mathbf{k}$ into an unpolarized hadron of transverse momentum $\mathbf{P}_{h\perp}$ with respect to the original quark direction. According to the Trento convention \cite{trento} the number density for finding an unpolarized hadron {\it h} produced from a transversely polarized quark {\it q} is defined as:
\begin{equation}
D_{h\/q^\uparrow}(z,P_{h\perp})  = D_1^q(z,P_{h\perp}^2) + H_1^{\perp q}(z,P_{h\perp}^2)\frac{(\hat{\mathbf{k}} \times \mathbf{P}_{h\perp})\cdot \mathbf{S}_q}{zM_h} ,
\label{eq:cdef}
\end{equation}
where the first term describes the unpolarized FF $D_1^q(z,P_{h\perp}^2)$, with $z\stackrel{CMS}{=} \frac{2E_h}{Q}$ being the fractional energy the hadron carries relative to half of the CMS energy Q. 
The second term, containing the Collins function $H_1^{\perp q}(z,P_{h\perp}^2)$, depends on the spin of the quark and thus, leads to an asymmetry as it changes sign if the quark spin flips. The vector product can accordingly be described by a $\sin(\phi)$ modulation, where $\phi$ is the azimuthal angle spanned by the transverse momentum and the plane defined by the quark spin and its momentum. 
In $e^+e^-$ hadron production the Collins effect can be observed when both a quark and an anti quark fragmentation are measured \cite{daniel}. Combining two hadrons from different hemispheres in jet-like events, with
azimuthal angles $\phi_1$ and $\phi_2$ as defined in Fig.~\ref{fig:angle2}, would result in a $\cos( \phi_1 + \phi_2 )$ modulation.
\section{Analysis}
\begin{figure}[t]
\begin{center}
\epsfig{figure=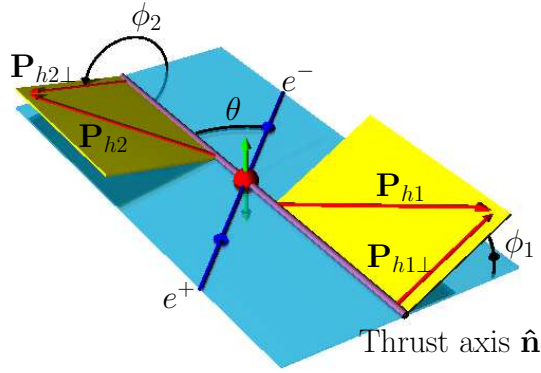,width=7cm}
\caption{Definition of the azimuthal angles of the two hadrons.
In each case, $\phi_i$ is the angle between the plane
spanned by the lepton momenta and the thrust axis $\hat{n}$, and the
plane spanned by $\hat{n}$ and the hadron transverse momentum $P_{h
i \perp}$.}
\label{fig:angle2}
\end{center}
\end{figure}

We select charged pion pairs within the central part of the detector by requesting them to have a polar angle $\theta_{1,2}$ relative to the beam axis $-0.6<\cos\theta_{1,2}<0.9$.  
One then measures the cosine modulation of the normalized pion pair yield in equ
idistant bins of the azimuthal angles $2\phi_0$ and $(\phi_1+\phi_2)$. $\phi_0$ is defined by the first hadron's transverse momentum relative to the 2$^{nd}$ hadron around the plane defined by the incoming lepton pair and the 2$^nd$ hadron. $\phi_{1,2}$ are defined around the thrust axis $\hat{n}$ (see Fig.~\ref{fig:angle2}), which is axproximately reflecting the original quark-antiquark axis. The thrust axis is obtained by maximizing the event shape variable thrust $t \stackrel{max}{=}\sum_h \frac{|\mathbf{P}_h \cdot \hat{\mathbf{n}}|}{|\mathbf{P}_h|}$, where over all detected particles exceeding a momentum of 0.1 GeV/c is summed. In order to only select two-jet like events a thrust value larger then 0.8 has been selected. 
Also the two pions have to be in opposing hemispheres by requiring $(\mathbf{P}_{h,1}\cdot \hat{\mathbf{n}})(\mathbf{P}_{h,2}\cdot \hat{\mathbf{n}}) <0$. To minimize hemisphere misidentification and the contribution from heavy quarks, the fractional energies $z\stackrel{CMS}{=}\frac{2E_h}{Q}$ to be larger than 0.2 and $Q_T$ has to be smaller than 3.5 GeV. 

\subsection{Double ratios}
The $\cos2\phi$ moments, where $2\phi$ stands for the two methods $2\phi_0$ and $\phi_1+\phi_2$, contain apart from the contribution of the Collins functions also radiative effects and possible also acceptance effects. Both of these are expected to be proportional to the unpolarized fragmentation functions. The unpolarized FFs appear also in the numerator of the cross sections due to normalization. As a consequence these additional terms are not sensitive to the charges the two pions carry and can thus be cancelled out by building double ratios of the normalized yields between unlike-sign and like-sign pairs. While the radiative termes cancel only to first order in the double ratios, they cancel exacly, when subtracting these two normalized yields from each other. However by subtracting on might not cancel the acceptance effects completely. The remaining, small differences between these two ways of cancelling the unwanted terms were assigned as systematic errors. Further systematic errors were obtained from the differences in the moments when introducing higher harmonics in the fit and from checking the double ratio method in MC which does not contain the Collins effect. Possible charge dependent effects were checked by building double ratios for positively charge pion pairs over negatively charged pion pairs. They were also assigned as systematic errors. Further systematic checks include the contribution from $e^+e^-\rightarrow \tau^+\tau^-$ events. Charm quark events were corrected by evaluating the asymmetries also in a charm enhanced data sample, where a $D^*$ candidate could be found. By reweighting MC with an additional azimuthal dependence the reconstruction could be tested. It was found that the results were underestimated in the $\cos(\phi_+\phi_2)$ method and had to be rescaled by a factor of 1.21. This can be explained by the average discrepancy between the thrust axis obtained from the generated events and the reconstructed events by 75mrad.
Also the correlations of the results have been tested and the statistical errors had to be rescaled by 14 \%.
 \section{Results}
\begin{figure}
\begin{center}
\epsfig{figure=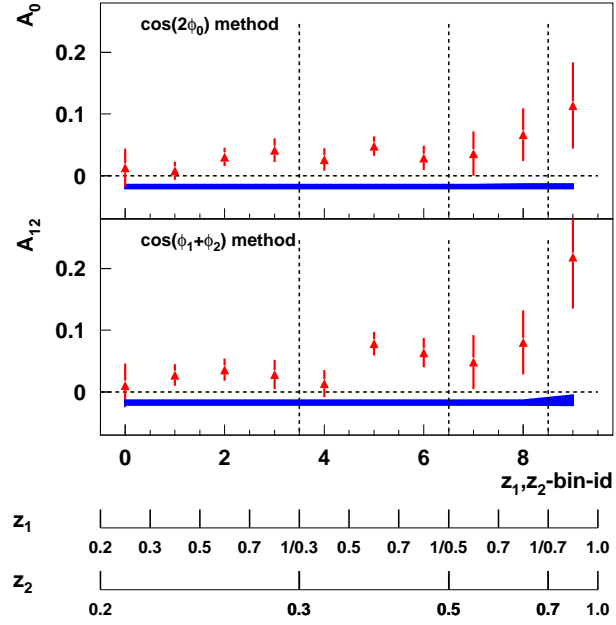,width=8cm}
\caption{Double ratios of unlike-sign over like-sign pion pairs versus combined z. The top plot shows the results of the $cos(2\phi_0)$ method, the bottom plot those of the $\cos(\phi_1+\phi_2)$ method.\label{fig:results}}
\end{center}
\end{figure}

The the $\cos2\phi$ results for the double ratios can be seen in Fig.~\ref{fig:results}, for a combined binning of the fractional energies $z_1$ and $z_2$. A clear rising behaviour can be seen and the results are large and nonzero. Further details can be found in \cite{prl}. 
These double ratio amplitudes contain the product of a quark and an antiquark Collins function. With the current double ratios however, it is not yet possible to clearly pin down the magnitude nor the functional dependence of the favored and disfavored Collins functions. Favored FFs describe the fragmentation of a quark into a hadron of same valence flavor, disfavored that of opposite valence flavor. As pointed out by \cite{peter}, the addition of charged pions as a third, independent information could help to pin down these questions. 

\section*{Acknowledgments}
The authors would like to thank D.~Boer for fruitful discussions on the theoretical aspects of the measurement. 
We thank the KEKB group for the excellent operation of the
accelerator, the KEK cryogenics group for the efficient
operation of the solenoid, and the KEK computer group and
the NII for valuable computing and Super-SINET network
support.  We acknowledge support from MEXT and JSPS (Japan);
ARC and DEST (Australia); NSFC (contract No.~10175071,
China); DST (India); the BK21 program of MOEHRD and the CHEP
SRC program of KOSEF (Korea); KBN (contract No.~2P03B 01324,
Poland); MIST (Russia); MHEST (Slovenia);  SNSF (Switzerland); NSC and MOE
(Taiwan); and DOE and NSF (USA).

\end{document}